\documentclass[useAMS,usenatbib]{mn2e}
\usepackage{epsfig}

\newcommand{\fei}{[Fe/H]$_{\rm I}$}
\newcommand{\feii}{[Fe/H]$_{\rm II}$}
\newcommand{\nafe}{[Na/Fe]}
\newcommand{\ofe}{[O/Fe]}
\newcommand{\nodata}{......}

\title[Variations in the Na-O anticorrelation]
{Variations in the Na-O anticorrelation in globular clusters:
Evidence for a deep mixing episode in red giant branch stars}
\author[Jae-Woo Lee]{Jae-Woo Lee$^{1}$\thanks{E-mail:jaewoolee@sejong.ac.kr}\\
$^{1}$Department of Astronomy and Space Science, ARCSEC, Sejong University,
Seoul 143-747, Korea}
\begin{document}

\date{Submitted 12 January 2010}

\pagerange{\pageref{firstpage}--\pageref{lastpage}} \pubyear{2010}

\maketitle

\label{firstpage}

\begin{abstract}
The Na-O anticorrelation seen in almost all globular clusters ever studied
using high-resolution spectroscopy is now generally explained by
the primordial pollution from the first generation of the intermediate-mass
asymptotic giant branch stars to the proto-stellar clouds of
the second generation of stars.
However, the primordial pollution scenario may not tell the whole story
for the observed Na-O anticorrelations in globular clusters.
Using the recent data by Carretta and his collaborators,
the different shapes of the Na-O anticorrelations for red giant branch stars
brighter than and fainter than the red giant branch bump can be clearly seen.
If the elemental abundance measurements by Carretta and his collaborators
are not greatly in error, this variation in the Na-O anticorrelation
against luminosity indicates an internal deep mixing episode
during the ascent of the low-mass red giant branch in globular clusters.
Our result implies that the multiple stellar population division scheme
solely based on [O/Fe] and [Na/Fe] ratios of a globular cluster,
which is becoming popular, is not reliable for stars
brighter than the red giant branch bump.
Our result also suggests that sodium supplied by the deep mixing
may alleviate the sodium under-production problem within
the primordial asymptotic giant branch pollution scenario.
\end{abstract}


\section{Introduction}
Variations in lighter elements, including oxygen and sodium, in globular cluster
(GC) red giant branch (RGB) stars have been known for more than three decades,
pioneered by \citet{cohen} and extensively studied by Lick-Texas group
\citep[see for example,][]{m15}.
It is very clear that anticorrelations between oxygen and sodium abundances
persist in almost all GCs ever studied using high-resolution spectroscopy
\citep{vii}.

The physical processes involved in the Na-O anticorrelation appear to be certain;
oxygen is depleted by the CNO cycle while sodium is enriched
from the $^{22}$Ne($p$, $\gamma$)$^{23}$Na reaction
in the hydrogen-burning shells of evolved stars.
The key issue has been the origin of this Na-O anticorrelation
and two viable scenarios has been debated;
the {\em evolutionary} and the {\em primordial} scenarios
\citep[see][]{k94}.

The evolutionary scenario is that a non canonical deep mixing in low-mass stars
during their ascent of RGB phase can supply freshly synthesized nuclides
to their photospheres \citep{langer}. The annoying trouble with the deep mixing
scenario is that Na-O anticorrelations does not appear to exist
in field halo stars with comparable metallicities to GCs.

The primordial scenario requires at least two generations of
star formation history in GCs.
Chemical pollution from the first generation intermediate-mass
asymptotic giant branch (AGB) stars \citep{ventura01} or
the fast rotating massive stars \citep{decressin} to the proto-stellar clouds
of the second generation of stars can produce the observed anticorrelations.
Although both the evolutionary and the primordial scenarios are required
to fully explain the detailed elemental abundance variations seen in GC stars,
consensus is in favor of the AGB pollution scenario
due to the presence of the Na-O anticorrelation in main sequence and
subgiant branch stars in GCs \citep{g01},
where the deep mixing can not be developed.
However, it should be noted that the AGB pollution scenario also
has a drawback that the amount of sodium synthesized in AGB stars
is not enough to explain the observed Na-O anticorrelations,
partly due to the uncertain cross-section of
the $^{23}$Na($p$, $\alpha$)$^{23}$Ne reaction rate,
which can destruct sodium in the later stage of AGB stars \citep{ventura06}.
To explain the observed Na-O anticorrelations in GCs within the AGB pollution
scenario, either an ad-hoc assumption of
a lowering $^{23}$Na($p$, $\alpha$)$^{23}$Ne reaction rate \citep{ventura06}
or a fine tuning of the AGB mass spectrum \citep{ventura08} are required.

Recently, Carretta and his collaborators \citep{n2808,n6218,n6752,vii,n6441}
provided a spectroscopic study for an unprecedented sample of RGB stars in 19 GCs,
collected in a homogeneous way.
Their work confirmed that the Na-O anticorrelations exist in all 19 GCs.
Furthermore, they devised a scheme to distinguish multiple stellar populations
by using [O/Fe] and [Na/Fe] ratios on the assumption that
{\em the primordial AGB pollution is entirely responsible
for the observed Na-O anticorrelations in GCs}.

In this Letter, using data provided by Carretta and his collaborators,
we show that different Na-O anticorrelations in GCs can be clearly seen
for RGB stars brighter than and fainter than the RGB bump.
It is believed that this variation in the Na-O anticorrelations against
luminosity is a strong evidence of the deep mixing episode 
during their ascent of RGB phase of low-mass stars,
if the elemental abundance measurements by Carretta and his collaborators
are not greatly in error.
Therefore, the primordial AGB pollution scenario may not tell the whole story
for the observed Na-O anticorrelation in GCs and
{\em the evolutionary deep mixing scenario should be resurrected}.

\begin{table}
{\tiny
\caption{Slopes in \fei\ and \feii\ versus $K$ mag.}
\begin{tabular}{lccrrrrr}
\hline
\multicolumn{1}{c}{Name} &
\multicolumn{1}{c}{$\langle$\fei$\rangle$} &
\multicolumn{1}{c}{$\langle$\feii$\rangle$} &
\multicolumn{2}{c}{\fei} &
\multicolumn{1}{c}{} &
\multicolumn{2}{c}{\feii} \\
\cline{4-5}\cline{7-8}
\multicolumn{1}{c}{}     &
\multicolumn{1}{c}{}     &
\multicolumn{1}{c}{}     &
\multicolumn{1}{c}{$a_{\rm I}$}  & \multicolumn{1}{c}{$\sigma(a_{\rm I})$} &
\multicolumn{1}{c}{}     &
\multicolumn{1}{c}{$a_{\rm II}$}  & \multicolumn{1}{c}{$\sigma(a_{\rm II})$} \\
\hline
\multicolumn{8}{c}{\it Metal-rich GCs}\\
N6441 & $-$0.33 & $-$0.29 &    0.091 & 0.051  & & $-$0.222 & 0.165  \\
N6388 & $-$0.41 & $-$0.35 &    0.015 & 0.011  & & $-$0.027 & 0.029  \\
N0104 & $-$0.74 & $-$0.77 &    0.001 & 0.001  & &    0.001 & 0.002  \\
N6838 & $-$0.81 & $-$0.80 & $-$0.003 & 0.008  & & $-$0.010 & 0.016  \\
&&&&&&&\\
\multicolumn{8}{c}{\it Intermediate metallicity GCs}\\
N6171 & $-$1.06 & $-$1.05 &    0.009 & 0.008  & &    0.011 & 0.020  \\
N2808 & $-$1.10 & $-$1.16 & $-$0.013 & 0.010  & & $-$0.055 & 0.015  \\
N6121 & $-$1.20 & $-$1.20 & $-$0.002 & 0.002  & & $-$0.002 & 0.004  \\
N0288 & $-$1.22 & $-$1.23 &    0.007 & 0.003  & &    0.000 & 0.009  \\
N6218 & $-$1.31 & $-$1.35 &    0.010 & 0.003  & & $-$0.005 & 0.009  \\
N5904 & $-$1.35 & $-$1.35 &    0.000 & 0.001  & &    0.003 & 0.002  \\
N3201 & $-$1.49 & $-$1.40 &    0.008 & 0.004  & & $-$0.009 & 0.010  \\
N1904 & $-$1.54 & $-$1.48 & $-$0.001 & 0.003  & &    0.021 & 0.005  \\
N6254 & $-$1.56 & $-$1.56 &    0.020 & 0.004  & &    0.027 & 0.009  \\
N6752 & $-$1.56 & $-$1.48 &    0.011 & 0.004  & & $-$0.023 & 0.009  \\
&&&&&&&\\
\multicolumn{8}{c}{\it Metal-poor GCs}\\
N6809 & $-$1.97 & $-$1.93 &    0.010 & 0.003  & &    0.001 & 0.005  \\
N6397 & $-$1.99 & $-$1.98 &    0.003 & 0.002  & &    0.059 & 0.009  \\
N4590 & $-$2.23 & $-$1.85 &    0.010 & 0.004  & &    0.225 & 0.044  \\
N7078 & $-$2.34 & $-$2.35 & $-$0.001 & 0.002  & &    0.001 & 0.007  \\
N7099 & $-$2.35 & $-$2.29 & $-$0.002 & 0.003  & &    0.018 & 0.012  \\
\hline
\end{tabular}
\label{tab:slope}}
\end{table}

\section{New interpretation of the sodium-oxygen anticorrelation}
\subsection{Slopes in the plots of ${\rm [Fe/H]}_{\rm II}$ versus $K$}
The widely used local-thermodynamic equilibrium (LTE) analysis
depends on appropriate stellar atmosphere model grids and input parameters,
such as effective temperature, surface gravity and turbulent velocity.
The derivation of stellar elemental abundances
is not a trivial task even for nearby bright stars.
The recent study of \citet{chara} may highlight the current situation.
They showed that the interferometric effective temperatures
for nearby K giant stars do not agree with those from spectroscopic observations,
suggesting a missing source of opacities in stellar atmosphere models.
The situation would be even worse for fainter stars, such as those in GCs.

\begin{figure}
\includegraphics[scale=0.65]{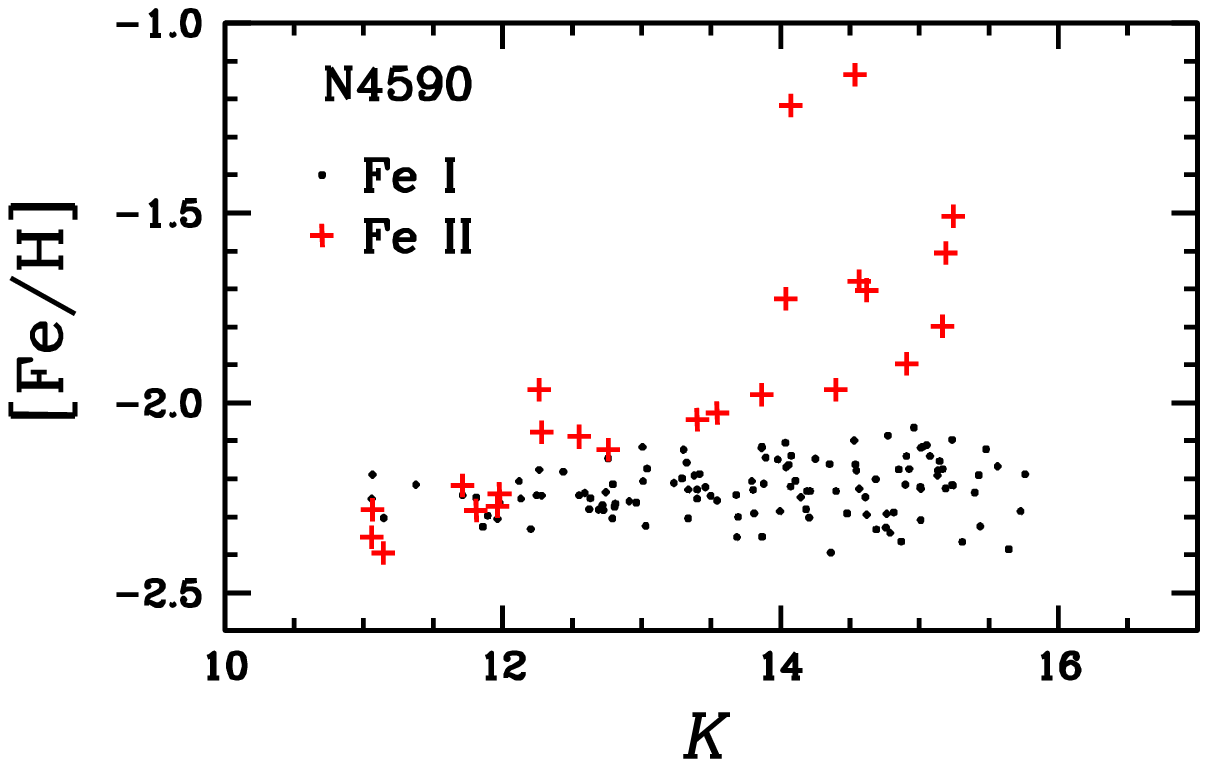}
\caption{A comparison of \fei\ and \feii\ against $K$ mag for RGB stars in NGC 4590
\citep{vii}. The slope in \feii\ versus $K$ may imply that
the surface gravities adopted by Carretta et al.\ are systematically in error.}
\label{fig:feh}
\end{figure}

The recent studies by Carretta and his collaborators
\citep{n2808,n6218,n6752,vii,n6441} provide \ofe\ and \nafe\ ratios
of RGB stars in 19 GCs, collected and analyzed in a homogeneous way.
Their oxygen abundances were based on the forbidden
lines of neutral oxygen at 6300.3 and 6363.8\AA,
which are very sensitive to adopted surface gravity,
and their \ofe\ ratios were derived from Fe I,
which is not very sensitive to surface gravity.
It is important to note that the \ofe\ ratio based on Fe II,
which is also very sensitive to surface gravity,
is more appropriate than that based on Fe I is \citep{ivans, m68}.

We examined the iron abundances of \citet[and references therein]{vii}
to see if their \feii\ ratios are consistent with \fei\ ratios,
which will validate their \ofe\ measurements based on Fe I.
We found a surprising discrepancy between \fei\ and \feii\
for many GCs.
Figure \ref{fig:feh} shows distributions of \fei\ and \feii\
against $K$ mag for RGB stars in NGC 4590,
as an extreme example\footnote{Since NGC 4590 is very metal-poor
(\fei\ = $-$2.24, see Table~\ref{tab:slope}),
Fe II absorption lines would be very weak.
However, line absorption measurement errors can not produce such a slope
seen in Figure~\ref{fig:feh}, unless their Fe II line measurements
are systematically in error with luminosity (i.e.\ S/N ratios of data).
The departure from the LTE can not explain this slope either,
since Fe I lines are more vulnerable to be affected than Fe II lines are.
Note that, in terms of \feii, NGC 4590 is not monometallic,
but has a metallicity spread of $>$ 1 dex.}. 
The figure clearly shows that the iron abundance derived from Fe II lines
has a substantial slope against $K$ mag, where $K$ mag of RGB stars
in a GC is related to the effective temperature and surface gravity
(i.e. a bright RGB star has lower effective temperature and
lower surface gravity; see for example \citealt{pal6}).
In Table \ref{tab:slope}, we show the slopes in \fei\ and \feii\
against $K$ mag for 19 GCs studied by \citet{vii}, assuming following relations,
\begin{eqnarray}
{\rm [Fe/H]}_{\rm I} &=& a_{\rm I}\times K +  b_{\rm I}, \nonumber\\
{\rm [Fe/H]}_{\rm II}&=& a_{\rm II}\times K +  b_{\rm II}. \nonumber
\end{eqnarray}
We also examined the variations in \fei\ and \feii\ against
effective temperatures of RGB stars in individual GCs.
Substantial slopes in the plots of \feii\ versus effective temperature
can be seen in many GCs, in sharp contrast to what \citet{vii} claimed
(see their Figure 5).

It is beyond the scope of our work to explain such discrepancy between
\fei\ and \feii\ in \citet{vii}.
However, as a first approximation, the gradient in \feii\ with respect to \fei\
may indicate that their adopted surface gravities are systematically in error.
For example, surface gravities of RGB stars in NGC 4590 could be
systematically larger with increasing $K$ mag.
The modification of surface gravity also tends to affect other input parameters,
such as effective temperature and turbulent velocity.
Therefore, the results presented by \citet{vii} may not be as accurate as
they claimed with {\em apparently} very small measurement errors.
Furthermore, their final results may not be as homogeneous as they claimed,
since the discrepancy between \fei\ and \feii\ varies one cluster
to the other and the break in the slope between \feii\ versus $K$ mag
or effective temperature can be seen in some GCs.
We note that if this discrepancy is entirely due to
incorrect surface gravity and other parameters adopted by \citet{vii},
their derivations of other elemental abundances, such as oxygen and sodium,
may not be reliable.
Also importantly, if the limited number of available Fe II lines
is responsible for this discrepancy \citep[see][]{n2808},
their [O/Fe] and [Na/Fe] ratios would be in the same
situation since only few lines are available for both elements.
We await their careful re-analysis of their data with keen interest.

\begin{figure}
\includegraphics[scale=0.4]{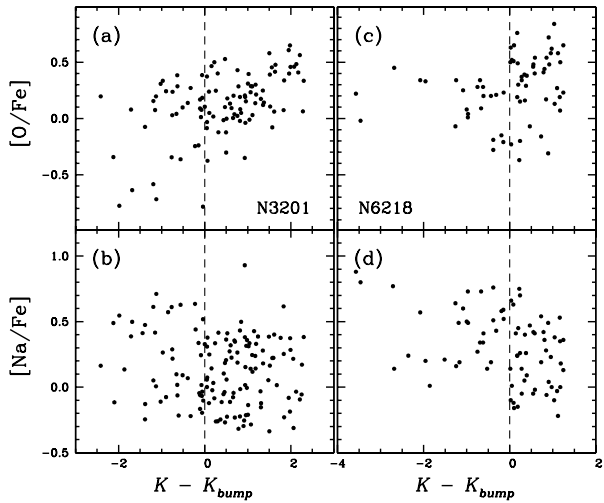}
\caption{Distributions of [O/Fe] and [Na/Fe] as functions of $K - K_{bump}$
for RGB stars in NGC 3201 and NGC 6218 \citep{n6218,vii}.
Note that both clusters do not show gradients in their \feii\
against $K$ mag, indicating that differential \ofe\ and \nafe\ ratios
may not be affected by potential systematic errors in surface gravity.
The differences in [O/Fe] and [Na/Fe] between
the upper RGB and the lower RGB stars can be clearly seen,
indicating that the deep internal mixing must have played some role
in the observed Na-O anticorrelation.}
\label{fig:n3201}
\end{figure}

\subsection{Variations in the Na-O anticorrelation against luminosity}
As we mentioned above, \citet{vii} derived [O/Fe] ratios based on Fe I and
a substantial slope in the \feii\ versus $K$ mag
may imply that their [O/Fe] ratios could be in error.
In Table~\ref{tab:slope}, eight GCs (NGC 6838, NGC 6121, NGC 288, NGC 6218,
NGC 5904, NGC 3201, NGC 6809 and NGC 7078) have negligibly small gradients
in their \fei\ and \feii\ versus $K$ mag.
Therefore, it is expected that {\em potential} systematic errors in
surface gravity adopted by \citet{vii} do not significantly affect
{\em differential} \ofe\ ratios against $K$ mag in these GCs.

In Figure~\ref{fig:n3201}, we show oxygen and sodium distributions
against $K-K_{bump}$ for RGB stars in NGC 3201 and NGC 6218\footnote{Although
$K_{bump}$ values, the $K$ mag at the RGB bump, are not known for many GCs,
we prefer to use $K$ bandpass since  some of GCs studied by \citet{vii}
are reported to have differential reddening and $K$ bandpass is less vulnerable
to differential reddening effect. We calculated the $K_{bump}$ values
for individual GCs using the relation given by \citet{cho}.
As shown in Table~\ref{tab:nao}, our $K_{bump}$ values are in good agreement
with those of \citet{cho}. We calculated the $K$ magnitude differences for 6 GCs
in common and we obtained the magnitude difference of
$\Delta K_{bump}$ = 0.00 $\pm$ 0.11 mag.}.
As shown in the figure, differences in \ofe\ and \nafe\ distributions
between the upper RGB and the lower RGB stars are notable\footnote{
Note that the recent study by \citet{dantona07} predicted the deep extra mixing
in RGB stars with an extreme helium abundance ($Y$ $\sim$ 0.4).
However, the absence of extended horizontal branch stars and
main-sequence split indicates that NGC 3201 and NGC 6218
do not contain subpopulations with such high helium abundances.}.
The mean \ofe\ ratio for the lower RGB stars is larger than that for
upper RGB stars in NGC 3201 and NGC 6218, while
the mean \nafe\ ratio for the lower RGB stars is smaller
than that for upper RGB stars.
If the results by \citet{vii} are not greatly in error,
are these variations in [O/Fe] and [Na/Fe] ratios
between the lower and the upper RGB
stars a signature of the internal deep mixing episode, so that
oxygen is depleted by the CNO cycle while sodium is enriched
from the $^{22}$Ne($p$, $\gamma$)$^{23}$Na reaction
in upper RGB stars of NGC 3201 and NGC 6218?

Table~\ref{tab:nao} summarizes the mean \ofe\ and \nafe\ for 
the bright RGB (bRGB, $K-K_{bump}$ $\leq$ $-$0.1 mag) group and
the faint RGB (fRGB, $K-K_{bump}$ $\geq$ +0.1 mag) group for individual GCs.
The differences in $\langle$[O/Fe]$\rangle$ and $\langle$[Na/Fe]$\rangle$ ratios
between the bRGB and the fRGB groups persist for most GCs,
in particular for those with  sufficient number of
stars in each subgroup (n $\geq$ 10).
The differences in $\langle$\ofe$\rangle$ and $\langle$\nafe$\rangle$
between the bRGB and the fRGB groups for above mentioned 8 GCs 
are at 2.71$\sigma$ and 1.92$\sigma$, respectively,
suggesting that Na-O anticorrelations between the bRGB and the fRGB
are different.
In addition to differences in $\langle$[O/Fe]$\rangle$ and $\langle$[Na/Fe]$\rangle$,
the scatters in the bRGB group are slightly larger than those in the fRGB group,
which makes sense if the internal deep mixing is partly responsible for 
the Na-O anticorrelation in the bRGB group.
If the measurement errors were responsible for these scatters,
an opposite would be expected.

\begin{figure}
\includegraphics[scale=0.4]{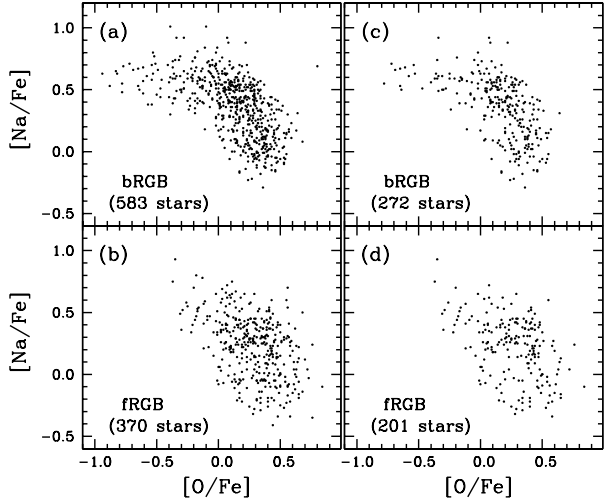}
\caption{(a -- b) A comparison of Na-O anticorrelations between 
the bRGB ($K-K_{bump}$ $\leq$ $-$0.1 mag, 583 stars) and 
the fRGB ($K-K_{bump}$ $\geq$ $-$0.1 mag, 370 stars) groups in 19 GCs.
(c -- d) As (a -- b) but for eight GCs 
(NGC 6838, NGC 6121, NGC 288, NGC 6218, NGC 5904, NGC 3201, NGC 6809 and NGC 7078)
with negligibly small gradients in \fei\ and \feii\ versus $K$ mag.
Lack of RGB stars with small [O/Fe] and large [Na/Fe] ratios
in the fRGB group is notable. This can not be explained by
the chemical pollution from the first generation AGB stars in GCs.}
\label{fig:nao}
\end{figure}

In Figure~\ref{fig:nao}, we show plots of [O/Fe] versus [Na/Fe] for
the bRGB and the fRGB groups in 19 GCs and in above mentioned 8 GCs.
It is very clear that the bRGB and the fRGB groups 
have different shapes of the Na-O anticorrelation.
We emphasize that the number of the E component with small \ofe\ and
large \nafe\ ratios \citep{vii}
in the fRGB group is smaller than that in the bRGB group.
If the Na-O anticorrelation observed in GCs are entirely due to
the previous generation of more massive stars,
the same number densities between the bRGB and the fRGB for
the second generation of stars (for example, the E component) would be expected.

\begin{figure}
\includegraphics[scale=0.4]{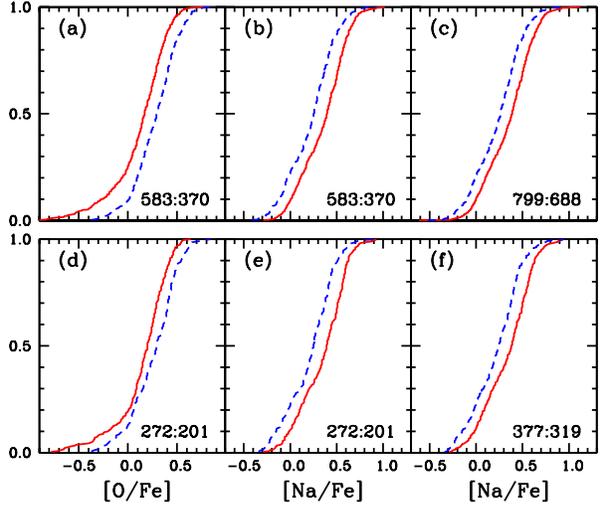}
\caption{(a \& b) Cumulative \ofe\ and \nafe\ distributions for the bRGB
(solid red lines, 583 stars)
and the fRGB (dashed blue lines, 370 stars) groups in 19 GCs.
The probabilities of being drawn from identical parent populations
are 0.00\% for both [O/Fe] and [Na/Fe], indicating that
the bRGB and the fRGB groups have different parent populations.
(c) Cumulative [Na/Fe] distributions
for all available stars in bRGB (799 stars) and fRGB (688 stars) groups.
(d \& e) As (a \& b) but for the bRGB (272 stars) and the fRGB (201 stars)
groups in 8 GCs.
The probabilities of being drawn from identical parent populations
are also 0.00\% for both \ofe\ and \nafe.
(f) As (c) but for the bRGB (377 stars) and the fRGB (319 stars) groups
in 8 GCs.}
\label{fig:dist}
\end{figure}

Cumulative distributions of \ofe\ and \nafe\ for the bRGB and the fRGB groups
for all 19 GCs and above mentioned 8 GCs
also show that they are really different as shown in Figure~\ref{fig:dist}.
We performed non-parametric Kolmogorov-Smirnov (K-S) tests to see if
the \ofe\ and the \nafe\ distributions of the two groups
are drawn from the same parent population.
Our calculations show that the probability of being drawn from
identical stellar populations is 0.00\% for both [O/Fe] and [Na/Fe],
strongly suggesting that they have different parent populations.
Perhaps, may difficulties involved in oxygen measurements for fRGB group
be responsible for the lack of the E component with depleted oxygen abundances?
Since the E component has large [Na/Fe] ratios,
the uncertainties in Na line measurements would be smaller.
Also note that \nafe\ ratio based on Fe I is much less sensitive
to surface gravity than \ofe\ ratio based on Fe I is.
Therefore potential systematic errors in surface gravity do not affect
the differential \nafe\ distributions.
We show cumulative \nafe\ distributions for all available stars
in the bRGB and the fRGB groups in Figure~\ref{fig:dist}-(c \& f)
and, again, they have very different distributions,
strongly suggesting that the internal deep mixing
occurred in bRGB stars is responsible for different
Na-O anticorrelations seen in the bRGB and the fRGB groups.

\begin{table*}
 \begin{minipage}{180mm}
{\tiny
\caption{Comparisons of $\langle$[O/Fe]$\rangle$ and $\langle$[Na/Fe]$\rangle$
between bRGB and fRGB.}
\begin{tabular}{lrrrrrrrrrrrrrrrrr}
\hline
\multicolumn{1}{c}{Name} &
\multicolumn{1}{c}{$K_{bump}$} &
\multicolumn{7}{c}{$\langle$[O/Fe]$\rangle$} &
\multicolumn{1}{c}{} &
\multicolumn{7}{c}{$\langle$[Na/Fe]$\rangle$} \\
\cline{3-9}\cline{11-17}\\
\multicolumn{2}{c}{} &
\multicolumn{3}{c}{bRGB} &
\multicolumn{1}{c}{} &
\multicolumn{3}{c}{fRGB} &
\multicolumn{1}{c}{} &
\multicolumn{3}{c}{bRGB} &
\multicolumn{1}{c}{} &
\multicolumn{3}{c}{fRGB}\\
\cline{3-5}\cline{7-9}\cline{11-13}\cline{15-17}\\
\multicolumn{2}{c}{} &
\multicolumn{1}{c}{n} &
\multicolumn{1}{c}{$\langle$[O/Fe]$\rangle$} &
\multicolumn{1}{c}{$\sigma\langle$[O/Fe]$\rangle$} &
\multicolumn{1}{c}{} &
\multicolumn{1}{c}{n} &
\multicolumn{1}{c}{$\langle$[O/Fe]$\rangle$} &
\multicolumn{1}{c}{$\sigma\langle$[O/Fe]$\rangle$} &
\multicolumn{1}{c}{} &
\multicolumn{1}{c}{n} &
\multicolumn{1}{c}{$\langle$[Na/Fe]$\rangle$} &
\multicolumn{1}{c}{$\sigma\langle$[Na/Fe]$\rangle$} &
\multicolumn{1}{c}{} &
\multicolumn{1}{c}{n} &
\multicolumn{1}{c}{$\langle$[Na/Fe]$\rangle$} &
\multicolumn{1}{c}{$\sigma\langle$[Na/Fe]$\rangle$} \\
\hline
\multicolumn{17}{c}{\it Metal-rich GCs}\\
N6441  &   14.51 &  21 &  0.041 &  0.168 &&   0 &\nodata &\nodata &&  21 &  0.392 &  0.287 &&   0 &\nodata &\nodata \\
N6388  &   14.15 &  25 & -0.039 &  0.255 &&   2 &  0.239 &  0.072 &&  29 &  0.404 &  0.242 &&   3 &  0.189 &  0.155 \\
N0104  &   12.03 &  83 &  0.139 &  0.186 &&   4 &  0.131 &  0.078 && 110 &  0.480 &  0.153 &&   6 &  0.494 &  0.108 \\
N6838  &   11.82 &  15 &  0.365 &  0.108 &&  10 &  0.400 &  0.099 &&  20 &  0.362 &  0.126 &&  11 &  0.337 &  0.135 \\
&&&&&&&&&&&&&&&&\\                                                
\multicolumn{17}{c}{\it Intermediate metallicity GCs}\\                
N6171  &   12.63 &   3 &  0.169 &  0.129 &&  23 &  0.198 &  0.174 &&   6 &  0.451 &  0.096 &&  26 &  0.314 &  0.208 \\
N2808  &   13.37 &  83 &  0.041 &  0.394 &&   0 &\nodata &\nodata && 112 &  0.271 &  0.257 &&   0 &\nodata &\nodata \\
N6121  &   10.13 &  55 &  0.193 &  0.116 &&  14 &  0.259 &  0.097 &&  74 &  0.342 &  0.205 &&  16 &  0.388 &  0.142 \\
N0288  &   13.16 &  24 &  0.075 &  0.301 &&  31 &  0.192 &  0.235 &&  37 &  0.340 &  0.275 &&  59 &  0.197 &  0.242 \\
N6218  &   11.78 &  24 &  0.150 &  0.226 &&  38 &  0.333 &  0.288 &&  31 &  0.443 &  0.233 &&  40 &  0.223 &  0.242 \\
N5904  &   12.75 &  69 &  0.131 &  0.306 &&  25 &  0.219 &  0.258 &&  89 &  0.312 &  0.246 &&  33 &  0.189 &  0.247 \\
N3201  &   11.80 &  28 & -0.036 &  0.340 &&  65 &  0.211 &  0.218 &&  37 &  0.208 &  0.293 &&  88 &  0.104 &  0.257 \\
N1904  &   13.74 &  21 &  0.016 &  0.231 &&  24 &  0.136 &  0.140 &&  26 &  0.410 &  0.227 &&  17 &  0.186 &  0.245 \\
N6254  &   11.39 &  24 &  0.133 &  0.285 &&  70 &  0.295 &  0.158 &&  33 &  0.239 &  0.239 &&  80 &  0.129 &  0.273 \\
N6752  &   11.15 &  37 &  0.218 &  0.283 &&  49 &  0.287 &  0.210 &&  48 &  0.425 &  0.254 &&  59 &  0.345 &  0.223 \\
&&&&&&&&&&&&&&&&\\                                                
\multicolumn{17}{c}{\it Metal-poor GCs}\\                                                                                       
N6809  &   11.61 &  52 &  0.202 &  0.212 &&  42 &  0.360 &  0.128 &&  64 &  0.378 &  0.305 &&  47 &  0.309 &  0.250 \\
N6397  &    9.73 &   3 &  0.252 &  0.118 &&   2 &  0.236 &  0.062 &&  10 &  0.084 &  0.335 &&  93 &  0.194 &  0.203 \\
N4590  &   12.87 &  19 &  0.280 &  0.185 &&  24 &  0.480 &  0.149 &&  19 &  0.263 &  0.174 &&  58 &  0.320 &  0.212 \\
N7078  &   12.86 &  21 &  0.188 &  0.180 &&  14 &  0.422 &  0.091 &&  25 &  0.367 &  0.256 &&  25 &  0.372 &  0.214 \\
N7099  &   12.40 &   8 &  0.027 &  0.291 &&  16 &  0.454 &  0.144 &&   8 &  0.416 &  0.270 &&  27 &  0.403 &  0.223 \\
\hline
Mean  &       &    &  0.134 & 0.227 &&    &  0.285$^1$ & 0.153$^1$ &&     &  0.347 & 0.235 &&     & 0.276$^1$ & 0.211$^1$ \\
      &       &    &$\pm$0.026&$\pm$0.020&&&$\pm$0.026$^1$&$\pm$0.017$^1$&&&$\pm$0.024&$\pm$0.015&&&$\pm$0.026$^1$&$\pm$0.012$^1$ \\
\hline
Mean$^2$&     &    &  0.160 & 0.231 &&    &  0.299 & 0.173 &&     &  0.341 & 0.236 &&     & 0.258 & 0.223 \\
      &       &    &$\pm$0.031&$\pm$0.022&&&$\pm$0.030&$\pm$0.019&&&$\pm$0.021&$\pm$0.014&&&$\pm$0.028&$\pm$0.012\\
\hline
Mean$^3$&     &    &  0.129 & 0.240 &&    &  0.285 & 0.188 &&     &  0.341 & 0.259 &&     & 0.255 & 0.228 \\
      &       &    &$\pm$0.032&$\pm$0.030&&&$\pm$0.033&$\pm$0.031&&&$\pm$0.027&$\pm$0.013&&&$\pm$0.040&$\pm$0.015\\
\hline
\multicolumn{17}{l}{$^1$17 GCs (without NGC 6441 and NGC 2808)}\\
\multicolumn{17}{l}{$^2$12 GCs (without NGC 6441, NGC 6388, NGC 104, NGC 6171, NGC 2808, NGC 6397 and NGC 7099)}\\
\multicolumn{17}{l}{$^3$8 GCs (NGC 6838, NGC 6121, NGC 288, NGC 6218, NGC 5904, NGC 3201, NGC 6809 and NGC 7078)}\\
\end{tabular}
\label{tab:nao}}
\end{minipage}
\end{table*}

\section{Summary}
We have shown that there exist variations in the Na-O anticorrelation against
the luminosity of RGB stars in GCs.
These variations can not be explained by chemical pollution from
the first generation intermediate-mass AGB stars in GCs and
are most probably due to the internal deep mixing during the ascent of RGB phase,
unless oxygen and sodium abundance measurements by \citet{vii}
are not greatly in error.

If the internal deep mixing can supply freshly synthesized sodium,
via the $^{22}$Ne($p$, $\gamma$)$^{23}$Na reaction,
to the photospheres of low-mass RGB stars in these GCs,
the trouble with the intermediate-mass AGB pollution scenario,
which requires more sodium production rate, can be alleviated.
However, the absence of the Na-O anticorrelation in the field halo stars
is still an open question.

Although, ejecta from the first generation of the intermediate-mass AGB stars
can significantly change primordial oxygen and sodium abundances
of the second generation of stars in GCs,
the multiple stellar population division scheme devised by \citet{vii},
which is solely based on [O/Fe] and [Na/Fe] ratios of individual GCs,
is not correct for bright RGB stars.
It is because when the internal deep mixing in low-mass RGB stars in GCs
is involved, one needs to disentangle the internal deep mixing effect
from the primordial intermediate-mass AGB contribution
to the observed Na-O anticorrelations in GCs.
An alternative approach by using 
additional photometric information, such as the $hk$ index which is known
to be not affected by the variations in lighter elemental abundances,
is more appropriate to distinguish the multiple stellar populations in GCs,
as \citet{nature} have discussed in detail.

\section*{Acknowledgments}
This work was supported by the faculty research fund of Sejong University in 2008.
Support for this work was also provided by the National Research Foundation of
Korea to the Astrophysical Research Center for the Structure and Evolution
of the Cosmos (ARCSEC).

\label{lastpage}

\end{document}